\documentclass[preprint,12pt]{elsarticle}

\usepackage[T1]{fontenc} 
\usepackage{subfig}
\usepackage{color}

\usepackage{amssymb}
\usepackage{amsmath}

\newcommand{\Dsl}{{\slash \negthinspace \negthinspace \negthinspace \negthinspace  D}}

\begin{document}

\begin{frontmatter}



\title{\boldmath Darkflation -- one scalar to rule them all?}

\author[adr1]{Zygmunt Lalak}
\ead{Zygmunt.Lalak@fuw.edu.pl}

\author[adr1]{and {\L}ukasz Nakonieczny\corref{cor1}}
\ead{Lukasz.Nakonieczny@fuw.edu.pl}
\cortext[cor1]{Corresponding author}

\address[adr1]{Institute of Theoretical Physics, Faculty of Physics, University of Warsaw \protect \\
ul.~Pasteura 5,~02-093 Warszawa, Poland }


\begin{abstract}
The problem of explaining 
both inflationary and dark matter physics in the framework of a minimal extension of 
the Standard Model was investigated. To this end, the Standard Model completed by a real
scalar singlet playing a role of the dark matter candidate has been considered. We assumed both the dark matter 
field and the Higgs doublet to be nonminimally coupled to gravity. 
Using quantum field theory in curved spacetime we 
derived an effective action for the inflationary period and analyzed its consequences.
In this approach, after integrating out both dark matter and Standard Model sectors 
we obtained the effective action expressed purely in terms of the gravitational field.
We paid special attention to determination, by explicit calculations, of the form of coefficients 
controlling the higher-order in curvature gravitational terms. Their connection to
the Standard Model coupling constants has been discussed. 
\end{abstract}

\begin{keyword}
inflation \sep quantum field theory in curved spacetime 



\end{keyword}

\end{frontmatter}



\section{Introduction}
Two of the most prominent challenges for modern cosmology are explanations of an exact nature of the inflationary period of the 
Universe history and the dark matter sector. By their exact nature we mean their realization in the context of field theory
and its connection to the Standard Model (SM) and gravitational physics. 
The dominant point of view in the researches on these subjects is that inflation and dark matter are two unrelated phenomena. 
The current article presents a different approach. As far as the dark matter problem is concerned, there is a consensus 
that a new matter field added to the SM is necessary in order to account for its presence. Using the minimalistic approach we chose to model the dark matter 
sector with the help of a real scalar singlet interacting with the SM via the Higgs portal \cite{Guo_2010,Cline_Scott_Kainulainen_Wengier_2013,Feng_2015}. 
Despite the fact that this model is very restricted in the low-mass region (the energies up to GeV scale), it still provides
a good dark matter candidate of the mass within the TeV range \cite{Feng_2015}. Although these masses are outside of the energy range available in the 
present direct detection experiments, this is the same range as proposed in many popular supersymmetric models, see for example 
\cite{Low_Wang_2014,Buchmueller2014,PhysRevD.92.035018,Roszkowski2015,PhysRevD.91.055011}.    

Another issue is the potential need of another new field to explain inflation. 
An agreement of the Starobinsky model of inflation \cite{Starobinsky_1980} with the recent Planck data 
\cite{Planck_20_inflation_2015} seems to suggest that the new field is not necessary. In the wake of the successful predictions of this model a large body of literature 
appeared on this subject. Most of it is concerned with the generalization of the gravitational action by including 
additional terms proportional to the Ricci scalar 
\cite{PhysRevD.90.023525,BenDayan_Jing_Torabian_Westphal_Zarate_2014,Sadeghi_2015,Artymowski_Lalak_Lewicki_2016,Artymowski2016432}. The authors recognized that the coefficients of
these terms, as coming from the loop corrections, should be connected to the coupling constants of the matter fields, yet they usually 
fixed their values on the ground of phenomenological considerations (see, e.g., \cite{BenDayan_Jing_Torabian_Westphal_Zarate_2014}).

In the current article, the calculations revealing exactly which couplings play a dominant role in defining the aforementioned coefficients 
in the considered setup were presented for the first time, to the best knowledge of the authors.
As a tool to obtain these results the heat kernel approach to quantum field theory in curved spacetime was employed \cite{Parker_Toms_2009,Buchbinder_Odintsov_Shapiro_1992,Avramidi_2000}.
Specifically, the procedure that was followed is mostly often used while investigating corrections coming from the quantum field representing 
matter to the renormalized energy-momentum tensor in curved spacetime 
\cite{Frolov_Zelnikov_1982,Frolov_Zelnikov_1984,Taylor_Hiscock_Anderson_2000,Matyjasek_2000,Matyjasek_Sadurski_2013}. The noninteracting fields are usually thus handled.
As far as interacting fields in curved spacetime are concerned, this methodology was used in finding quantum corrections 
to the tree-level effective action (in the context of the running of the couplings) 
\cite{Elizalde_Odintsov_1994,Elizalde_Odintsov_1994_2,Elizalde_Kirsten_Odintsov_1994,Elizalde_Odintsov_Romeo_1995}. 
An extension of this idea to the inflationary physics was presented in \cite{PhysRevD.90.084001,Salvio2015194,PhysRevD.91.083529}, especially the latter two papers deals with the running of couplings constants
in the gravitational sector.  

More recently, it was used in constructing the one-loop 
corrected effective action for various matter fields in nontrivial gravitational background (e.g., curvaton-inflaton dynamic \cite{Markkanen_Tranberg_2012})
and during examining the stability of the Higgs effective potential 
\cite{Herranen_Markkanen_Nurmi_Rajantie_2014,Herranen_Markkanen_Nurmi_Rajantie_2015,Czerwinska_Lalak_Nakonieczny_2015}.

At this point we want to stress that this minimalistic approach to explaining inflation and dark matter in a single consistent framework
has been already used with some success at the tree-level where the coefficient of the $R^2$ term was fixed by phenomenology, see, e.g., \cite{Gorbunov2011157,Gorbunov201215,Calmet_Kuntz_2016}.
What we investigated in this article was the following problem: 
can we neglect the $R^2$ term at the tree-level and instead rely on the loop effects to generate it? 
In the process we derived functional forms of the terms higher-order in curvature arising at the one-loop level and found the 
connection between their coefficients and the matter coupling constants.

Another important remark is that although the idea of realizing both dark matter and inflation by a single scalar field already has been discussed in the literature (see for example 
\cite{PhysRevD.80.123507}), our approach is conceptually different. We do not postulate that an additional scalar should be both the dark matter candidate and the inflaton.
Instead, using a tool of the quantum field theory in curved spacetime, we integrate out both dark matter and Standard Model sectors and end up with the effective action 
given solely in terms of the gravitational scalars, namely the Ricci scalar, the Kretschmann scalar and a square of the Ricci tensor.

The structure of the article is the following. In section \ref{sec:intro} we present and discuss the employed model of the matter and
gravity sectors. Then we focus on parts that are important for obtaining the one-loop effective action for the problem at hand.
In section \ref{sec:inflation} we obtain and analyze properties of the effective field theory 
relevant for the inflationary period of the history of the Universe. 
The last section contains a summary of the obtained results.

\section{General form of the action}
\label{sec:intro}

We start our investigations by specifying the action for the gravity and matter sectors, its tree-level unrenormalized form is given by 
\begin{align}
S_g = \int \sqrt{-g}\ d^4 x &\left [  - \frac{1}{16 \pi G} \left ( R + 2 \Lambda \right ) + 
\alpha_1 R_{\alpha \beta \mu \nu}R^{\alpha \beta \mu \nu} + \right. \nonumber \\
&+ \alpha_2 R_{\alpha \beta}R^{\alpha \beta} +\alpha_3 R^2   + 
\alpha_4 \square R\Big],  \\
\label{S_matter}
S_m = \int \sqrt{-g}\ d^4 x\, &\bigg \{  \left [ d_{\mu} \tilde{X}\right ] d^{\mu} \tilde{X} - m^2_{X} \tilde{X}^2 + \xi_{X} \tilde{X}^2 R - 
\tilde{m}_0 \tilde{X} - \lambda_X \tilde{X}^4 +  \nonumber \\
&~ - \tilde{m}_{1} \tilde{X} R  - \tilde{m}_{3} \tilde{X}^3 +  \nonumber \\
&~+  \left [ d_{\mu} \tilde{H} \right ]^{\dagger} d^{\mu} \tilde{H} - m_{H}^2 |\tilde{H}|^2 - \frac{\lambda_{H}}{2}|\tilde{H}|^4 + 
\xi_H |\tilde{H}|^2 R +\nonumber \\
&~ - \tilde{m}_{HX} \tilde{X} |\tilde{H}|^2 - \lambda_{HX} \tilde{X}^2 |\tilde{H}|^2 +  \nonumber \\
&~+ \bar{\psi}_{Q} i \Dsl P_{+} \psi_{Q} + \bar{\psi}_{U} i \Dsl P_{-} \psi_{U} + \nonumber \\
&~- y \bar{\psi}_{Q} P_{-} \left [ i \sigma^2 \tilde{H}^{*} \right ] \psi_{U} - 
y \bar{\psi}_{U} P_{+} \left [ - i \tilde{H}^{T} \sigma^2 \right ] \psi_{Q} \bigg \}.
\end{align}
The field content of the matter part of the action is $\tilde{X}$ -- an additional real scalar singlet, 
$\tilde{H}$ -- the complex Higgs doublet, $\psi_{Q}$ is the left-handed quark doublet extended to the form the full
Dirac spinor, namely $\psi_{Q} = \begin{bmatrix} Q \\ Q\end{bmatrix}$, where $Q$ is the usual Standard Model doublet, 
$\psi_{U}$ is the Dirac spinor created from the SM right handed singlet in a similar manner. 
This formal extension of the fermionic sector is necessary in order to apply the heat kernel method to this sector,
namely during calculations of traces of the Schwinger-DeWitt coefficients the properties of the Clifford algebra of the four 
dimensional gamma matrices are used extensively, see for example \cite{Parker_Toms_2009}. To keep the correct number of the 
fermionic degrees of freedom we used the standard projection operators $P_{\pm} \equiv \frac{1 \pm \gamma^5}{2}$, where 
$\gamma^5 \equiv i \gamma^0 \gamma^1 \gamma^2 \gamma^3$.     
Since we do not expect nontrivial gauge background during inflationary period we may for now disregard 
the gauge boson sector of the Standard Model.
Under this assumption the symbols $d_{\mu}$ and $\Dsl$ reduce to the covariant derivative and the covariant Dirac operator, namely 
\begin{align}
d_{\mu} \equiv \nabla_{\mu} ,\quad 
\Dsl \equiv \gamma^{\mu} d_{\mu} .
\end{align}
In the next step we divide fields into the background part and the quantum fluctuation
\begin{align}
\label{fields_split}
\tilde{X} &= \hat{X} + X, \nonumber \\
\tilde{H}^{(p)} &= \hat{H}^{(p)} + H^{(p)}, \nonumber \\
\tilde{H}^{* (p)} &= \hat{H}^{* (p)} + H^{* (p)}, \\
\psi_{Q} &= \hat{\psi}_{Q} + \chi_{Q}, \nonumber  \\
\psi_{U} &= \hat{\psi}_{U} + \chi_{U}, \nonumber 
\end{align}
where a hat indicates the quantum part and quantities without it represent a classical background. From now on, we will also use
the following notation for the Higgs field: $\tilde{H}^{(p)} = \left[ \tilde{H}^{(1)}, \tilde{H}^{(2)} \right]^{T}$, where $\tilde{H}^{(1)}$ 
and $\tilde{H}^{(2)}$ are complex scalars. Below we present the part of the action that is quadratic in the quantum fluctuations 
and defines the Laplace-Beltrami operator that is crucial in the heat kernel approach for constructing the effective action \cite{Parker_Toms_2009,Buchbinder_Odintsov_Shapiro_1992,Avramidi_2000}.
To obtain it, we put relations (\ref{fields_split}) into the action (\ref{S_matter}) and keep only terms that are quadratic in quantum fields. 
Moreover, for convenience, we split the total action quadratic in fluctuations into a few parts $S^{(2)}_{matter} = S_{scalar-scalar} + S_{fermion-fermion} +
S_{fermion-scalar} + S_{scalar-fermion}$, where:
\begin{align}
\label{S2_scalar}
S_{scalar-scalar} &= \int \sqrt{-g}\ d^4 x\; \bigg \{
\hat{X} \Big( - \square - m_X^{2} + \xi_X R - 6 \lambda_X X^2 - 
\lambda_{HX} |H|^2 \Big) \hat{X} + \nonumber \\
&+ \hat{H}^{* (p)} \Big[ - \square_{p q} - m_{H}^2 \delta_{ p q} + \xi_{H}R \delta_{p q} - \tilde{m}_{HX} X \delta_{p q} +  \nonumber \\
&- \frac{\lambda_{H}}{2} \Big( 2|H|^2 \delta_{p q} + 2 \delta_{p r} H^{(r)} H^{* (s)} \delta_{s q} \Big) - \lambda_{HX} X^2 \delta_{pq} \Big] \hat{H}^{(q)}  +  \nonumber \\
&+ \hat{H}^{* (p)} \bigg( - \frac{\lambda_{H}}{2} \delta_{p r} H^{(r)} \delta_{q s} H^{(s)} \bigg) \hat{H}^{* (q)} +  \nonumber \\
&+ \hat{H}^{(a)} \bigg( - \frac{\lambda_{H}}{2} H^{*(c)} \delta_{a c} H^{*(d)} \delta_{d b} \bigg) \hat{H}^{(b)} +  \nonumber \\
&+ \hat{X} \hat{H}^{* (p)} \Big( -  \tilde{m}_{HX} \delta_{p q} H^{(q)} - 2 \lambda_{HX} X \delta_{p q} H^{q} \Big) + 
 \nonumber \\
&+ \hat{X} \hat{H}^{(p)} \Big(- \tilde{m}_{HX} H^{*(q)} \delta_{p q }  - 2 \lambda_{HX} X H^{* q} \delta_{pq} \Big) 
\bigg \},
\end{align}
in the above formula $\square \equiv d_{\mu}d^{\mu}$ stands for the covariant d'Alembertian,
\begin{align}
S_{fermion-fermion} &= \int \sqrt{-g}\ d^4 x \left(
i \hat{\bar{\psi}}_{Q} \gamma^{\mu} d_{\mu } P_{+} \hat{\psi}_{Q} + i \hat{\bar{\psi}}_{U} \gamma^{\mu} d_{\mu} P_{-} \hat{\psi}_{U} + 
\right. \nonumber \\
&\left. - y \hat{\bar{\psi}}^{(p)}_{Q} P_{-} \epsilon_{p q} H^{* (q)} \hat{\psi}_{U} + 
y \hat{\bar{\psi}}_{U}  \epsilon_{p q}  H^{(p)} P_{+} \hat{\psi}^{(q)}_{Q} \right),\\
S_{fermion-scalar} &= \int \sqrt{-g}\ d^{4}x \left(
- y \hat{\bar{\psi}}^{(p)}_{Q} P_{-} \epsilon_{p q} \hat{H}^{*(q)} \psi_{U} + \right. \nonumber \\
&\left. + y \hat{\bar{\psi}}_{U} P_{+}  \epsilon_{p q} \hat{H}^{(p)} \psi^{(q)}_{Q} 
\right), \\
S_{scalar-fermion} &= \int \sqrt{-g}\ d^{4}x \left(
- y \bar{\psi}^{(p)}_{Q} P_{-} \epsilon_{p q} \hat{H}^{* (q)} \hat{\psi}_{U} + \right. \nonumber \\
&\left. 
+ y \bar{\psi}_{U} P_{+} \epsilon_{p q} \hat{H}^{(p)} \hat{\psi}^{(q)}_{Q}
\right).
\end{align}
The first step in using the heat kernel method is to rewrite this quadratic part in the form 
that makes reading off the appropriate matrix form of Laplace-Beltrami operator easy, namely 
\begin{align}
\label{S2_general}
S_{matter}^{(2)} = - \int \sqrt{-g}\ d^4 x\; \hat{\Phi}^{* T} D^2 \hat{\Phi},
\end{align} 
where $\hat{\Phi} = \Big[ \hat{X}, \hat{H}^{(a)}, \hat{H}^{* (p)}, \hat{\psi}^{(p)}_{Q}, \hat{\psi}_{U} \Big]^{T}$ and 
the differential operator $D^2$ is of the form
\begin{align}
D^2 = \square + 2 h^{\mu} d_{\mu} + \Pi.
\end{align}
The symbol ${}^T$ in (\ref{S2_general}) stands for the operation that transforms the multiplet $\Phi$ represented by the column vector to 
the same multiplet represented by the row vector. This operation does not transform $\psi$ into $\psi^{T}$. On the other hand, 
${}^{*}$ stands for the complex conjugate of the field $H^{p}$ and the Dirac conjugate for the spinor field.  
To obtain the aforementioned simple form of the fluctuation action we redefine the quantum fields in the following way: 
\begin{align}
\hat{\Phi} \rightarrow \Big[ 
 \hat{X}, \sqrt{2} \hat{H}^{(p)},  \sqrt{2} \hat{H}^{* (p)},  i \gamma^{\nu}d_{\nu} \hat{\psi}^{(p)}_{Q}, 
i \gamma^{\nu}d_{\nu} \hat{\psi}_{U} \Big]^{T}.
\end{align}
This redefinition has twofold consequences. The first one is that the redefinition of the quantum fields brings the Jacobian factor in the 
path integral. For the scalar field this is an irrelevant number. For the fermionic one this leads only to an 
appearance of the purely gravitational terms that could be absorbed into the tree-level action by redefining the
renormalized gravitational constant and $\alpha_i$ constants in the front of the terms of higher order in curvature \cite{Czerwinska_Lalak_Nakonieczny_2015}. 
The second consequence of this redefinition is the transformation of the fluctuation part of the action into the form
\begin{align}
\label{S2_scalar_t}
S_{scalar-scalar} &=  - \int \sqrt{-g}\ d^4 x\; \bigg \{ \hat{X} \Big(  \square + m_X^{2} - \xi_X R  
 + 6 \lambda_{X} X^2 + \lambda_{HX} |H|^2 \Big) \hat{X} + \nonumber \\
& + 2 \hat{H}^{* (p)} \Big[ \Big( \square + m_{H}^2  - \xi_{H}R  + \lambda_{H} |H|^2 +
 \tilde{m}_{HX} X + \lambda_{HX} X^2 \Big) \delta_{p q} +  \nonumber \\
& + \lambda_H \delta_{p r} H^{(r)} H^{* (s)} \delta_{s q} 
 \Big] \hat{H}^{(q)}  
+ \hat{H}^{* (p)} \Big( \lambda_{H} \delta_{p r} H^{(r)} \delta_{s q} H^{(s)} \Big) \hat{H}^{* (q)} + \nonumber \\
& + \hat{H}^{(p)} \Big(  \lambda_{H} H^{*(s)} \delta_{p s} H^{*(r)} \delta_{r q} \Big) \hat{H}^{(q)} +  \nonumber \\
& + \sqrt{2} \hat{X} \hat{H}^{* (p)} \Big( \tilde{m}_{HX} \delta_{p q} H^{(q)} + 2 \lambda_{HX} X \delta_{p q} H^{q} \Big)  + \nonumber \\
& + \sqrt{2} \hat{X}^{*} \hat{H}^{(p)} \Big( \tilde{m}_{HX} H^{*(q)} \delta_{p q } + 2 \lambda_{HX} X \delta_{p q} H^{* q}\Big) 
\bigg \}, \\
\label{S2_fermion_t}
S_{fermion-fermion} &= - \int \sqrt{-g}\ d^4 x \left [
\hat{\bar{\psi}}_{Q} \left( \square - \frac{1}{4}R \right) P_{+} \hat{\psi}_{Q} +\right. \nonumber \\
&\left. + \hat{\bar{\psi}}_{U} \left( \square - \frac{1}{4}R \right) P_{-} \hat{\psi}_{U} + \right. \nonumber \\
&\left. 
+ i y \hat{\bar{\psi}}^{(p)}_{Q} P_{-} \epsilon_{p q} H^{* (q)} \gamma^{\mu} d_{\mu} \hat{\psi}_{U} +\right. \nonumber \\
&\left.- i y \hat{\bar{\psi}}_{U}  \epsilon_{p q} H^{(p)} P_{+} \gamma^{\mu} d_{\mu} \hat{\psi}^{(q)}_{Q} \right ], \\
\label{S2_fermion_scalar_t}
S_{fermion-scalar} &= - \int \sqrt{-g}\ d^{4}x \Big(
 \sqrt{2} y \hat{\bar{\psi}}^{(p)}_{Q} P_{-} \epsilon_{p q} \hat{H}^{*(q)} \psi_{U} + \nonumber \\
&-\sqrt{2} y \hat{\bar{\psi}}_{U} P_{+} \epsilon_{p q} \hat{H}^{(p)} \psi^{(q)}_{Q} \Big), \\
\label{S2_scalar_fermion_t}
S_{scalar-fermion} &= - \int \sqrt{-g}\ d^{4}x \Big(
i \sqrt{2} y  \bar{\psi}^{(p)}_{Q} P_{-} \epsilon_{p q} \hat{H}^{* (q)} \gamma^{\mu} d_{\mu} \hat{\psi}_{U} + \nonumber \\
&-i \sqrt{2} y \bar{\psi}_{U} P_{+}  \epsilon_{p q} \hat{H}^{(p)} \gamma^{\mu} d_{\mu} \hat{\psi}^{(q)}_{Q} \Big).
\end{align}
Having the above in mind, we may construct a matrix form of the operator we sought. Appropriate entries of the matrices 
$h^{\mu}$ and $\Pi$ could be read off from the $S_{sclalar-scalar}$, $S_{fermion-fermion}$, $S_{scalar-fermion}$ and $S_{fermion-scalar}$.
\footnote{For example, as is evident from the relations (\ref{S2_scalar_t})--(\ref{S2_scalar_fermion_t}), the only contributions to $h^{\mu}$
will come form (\ref{S2_scalar_fermion_t}), moreover the explicit matrix form of $h^{\mu}$ is given by
\begin{align}
2h^{\mu} = \begin{bmatrix}
0&0&0&0&0\\ 
0&0&0&0& i \sqrt{2} y  \bar{\psi}^{(p)}_{Q} P_{-} \epsilon_{p q}  \gamma^{\mu} \\
0&0&0& - i \sqrt{2} y \bar{\psi}_{U} P_{+}  \epsilon_{p q}  \gamma^{\mu} &0 \\
0&0&0&0&0\\ 
0&0&0&0&0
\end{bmatrix}.\nonumber 
\end{align}
}

\section{Effective action in the inflationary era}
\label{sec:inflation}

In this section we will focus on the inflationary era of the history of the Universe. Our goal is to check whether quantum corrections
coming form the presence of the scalars nonminimally coupled to gravity could indeed lead to the Starobinsky-like action
for the gravity sector. The second goal is to check if there is enough freedom in the choice of the parameters of the theory
to get good inflationary predictions. To simplify our discussion we will disregard fermionic contributions to the effective action.
Their possible presence will not introduce additional new types of terms in the gravity sector but will only lead to the numerical 
changes in the coefficients (which are subdominant compared to the coming from the scalar sector) \cite{Czerwinska_Lalak_Nakonieczny_2015}.   
Using the heat kernel approach and dimensional regularization we may express the unrenormalized one-loop part of the effective action as
\begin{align}
\label{S_one_loop_gen}
\Gamma^{(1)} &= \frac{ i \hbar}{2} \ln \text{Det}{\mu^{-2}D^2} = \nonumber \\
& = \hbar \int \sqrt{-g}\ d^4 x\; \frac{1}{64 \pi^2} \text{Tr} \bigg \{
\tilde{a}_{0} M^4 \bigg [ \frac{2}{\bar{\varepsilon}} 
- \ln \bigg ( \frac{M^2}{\mu^2} \bigg )  + \frac{3}{2} \bigg ] + \nonumber \\
&- 2 \tilde{a}_{1} M^2 \bigg [ 
\frac{2}{\bar{\varepsilon}} + 1 - \ln \left ( \frac{M^2}{\mu^2} \right )
\bigg ] + 2 \tilde{a}_{2} \bigg [ \frac{2}{\bar{\varepsilon}} - \ln \bigg ( \frac{M^2}{\mu^2} \bigg )  \bigg ]
+ M^{-2} \tilde{a}_{3}
\bigg \},
\end{align}
where $\frac{2}{\bar{\varepsilon}} = \frac{2}{\varepsilon} - \gamma + \ln(4\pi)$, $\gamma$ is the Euler constant, $n = 4 - \varepsilon$
and $\text{Tr}$ stands for the matrix trace and sum over all discrete indices (group or Lorentz ones).
The R-summed Schwinger-DeWitt coefficients are given by \cite{Parker_Toms_1985,Jack_Parker_1985}
\begin{align}
\tilde{a}_{0} &= 1, \nonumber \\
\tilde{a}_{1} &= 0, \\
\tilde{a}_{2} &= \frac{1}{180} \Big(
-  R_{\mu \nu} R^{\mu \nu} +  R_{\alpha \beta \mu \nu} R^{\alpha \beta \mu \nu} + \square R
 \Big) {\bf{1}} + \frac{1}{6} \square M^2  + \frac{1}{12} W_{\alpha \beta} W^{\alpha \beta}, \nonumber
\end{align} 
where $\bf{1}$ is the unit matrix in the field space and there is the following relation between the differential operator $D^2$ and matrices $M^2$ and $W_{\alpha \beta}$:
\begin{align}
D^2 &= \square {\bf{1}} + 2 h^{\mu} d_{\mu} + \Pi, \\
\label{M2def}
M^2 &= \Pi + \frac{1}{6}R {\bf{1}} - d_{\mu} h^{\mu} -h_{\mu} h^{\mu}, \\
W_{\alpha \beta} &= \left [ d_{\alpha}, d_{\beta} \right ] {\bf{1}}  + 2 d_{[ \alpha} h_{ \beta ]} + 
\left [ h_{\alpha} , h_{\beta}\right ].
\end{align}
In the above formulas, $d_{\mu}$ is a covariant derivative containing both gravity and gauge (if present) parts and $\square  = d_{\mu} d^{\mu}$.
Using the $\overline{\textrm{MS}}$ (modified Minimal Subtraction, for introduction see \cite{Peskin_Schroeder_1995}) renormalization scheme 
we may write the leading log part of the one-loop effective action in the large mass expansion as 
\begin{align}
\label{G1_gen}
\Gamma^{(1)} = \frac{1}{64 \pi^2} \int \sqrt{-g}\ d^4 x  &\text{Tr} \bigg \{ M^4 \bigg [ - \ln \bigg (\frac{M^2}{\mu^2} \bigg ) + \frac{3}{2} \bigg ] - 2 \tilde{a}_{2} \ln \bigg (\frac{M^2}{\mu^2} \bigg ) \bigg \}.
\end{align}
It is worthy to note that this leading log approximation is valid as long as $\frac{\mathcal{R}^3}{M^2} < \mathcal{R}^2$,
where $\mathcal{R}^3$ ($\mathcal{R}^2$) represents all terms that are of the third (second) order in curvature. For our setup
this is a good approximation (details will be given shortly) for the inflationary period provided that inflation ends in the 
de Sitter or dust dominated stage. This assumption means that the transition to the radiation dominated era happens 
during the reheating period of the cosmological history.

\subsection{Higgs and dark matter sector during the inflationary epoch}

As was mentioned at the beginning of this section, to model the inflation period of the history of the Universe we will use the Starobinsky-like approach.
To this end, we will consider a second scalar to possess a large mass parameter $m_{X}^2 > 0$ and
a linear and possibly quartic coupling to the Higgs doublet.
The rest of the parameters present in the scalar part of (\ref{S_matter}) will be given 
by $\tilde{m}_{1} = \tilde{m}_{2} = \tilde{m}_{3} = \lambda_{X} = 0$. This gives us the following form of the
tree-level potential at the beginning of inflation:
\begin{align}
V &= \frac{1}{2} m_{X}^2 Y^2 - \frac{1}{2} \xi_{X} R Y^2 + \frac{\tilde{m}_{HX}}{2 \sqrt{2}} Y h^2 + \frac{1}{4} \lambda_{HX} h^2 Y^2 + \nonumber \\ 
&- \frac{1}{2} |m_{H}^2|  h^2 + \frac{\lambda_{H}}{8} h^4 - \frac{1}{2} \xi_{H} R h^2,
\end{align}
where we used the rescaling $X = \frac{1}{\sqrt{2}} Y$ and $H^{(2)} = H^{* (2)}= \frac{1}{\sqrt{2}} h$.
Looking for minima of this potential we found out that for $\xi_{H}>0$ and $\xi_{X}>0$ we have one minimum with $Y = 0$ 
and $h = 0$. This is true as long as $ - \xi_{H} R - |m_{H}^2| >0$.
Below, the contributions to the effective action coming from the scalar sector are presented. 
They come in two parts, namely the scalar-Higgs and the Goldstone boson ones.
The scalar-Higgs contribution is given by
\begin{align}
m_{1}^{Yh} = \begin{bmatrix}
 \tilde{M}_{YY}^2 & \frac{ \tilde{m}_{HX}}{ 2 } h + \frac{\lambda_{HX}}{2} Yh & \frac{ \tilde{m}_{HX}}{ 2 } h + \frac{\lambda_{HX}}{2} Yh\\
 \frac{ \tilde{m}_{HX}}{ 2 } h + \frac{\lambda_{HX}}{2} Yh & \tilde{M}_{hh}^2  & \frac{1}{2} \lambda_H h^2 \\
 \frac{ \tilde{m}_{HX}}{ 2 } h + \frac{\lambda_{HX}}{2} Yh & \frac{1}{2} \lambda_{H} h^2 & \tilde{M}_{hh}^2
\end{bmatrix},
\end{align} 
where 
\begin{align}
\tilde{M}_{YY}^2 &= \left (  \frac{1}{6} - \xi_X \right ) R + m_{X}^2  + \frac{\lambda_{HX}}{2} h^2 + 3 \lambda_X Y^2 , \\
\tilde{M}_{hh}^2 &= \left (  \frac{1}{6} - \xi_H \right ) R + m_{H}^2 +  \lambda_{H} h^2 + \frac{\lambda_{HX}}{2} Y^2 + \frac{ \tilde{m}_{HX}}{\sqrt{2}} Y.
\end{align}
The matrix $m^{Yh}_{1}$ encompasses contributions to the $M^2$ matrix defined in (\ref{M2def}) from the dark matter candidate $Y$, the real neutral component of the Higgs field $h$ 
and bosons described by the imaginary parts of $H^{(2)}$.
The remaining two Goldstone bosons contribute as 
\begin{align}
m^{G}_{1} = \begin{bmatrix}
\tilde{M}_{G}^2  & 0 \\
0 & \tilde{M}_{G}^2 
\end{bmatrix},
\end{align}
where $\tilde{M}_{G}^2 = m_H^2 + \left ( \frac{1}{6} - \xi_H \right ) R + \frac{1}{2} \lambda_H h^2 + \frac{\lambda_{HX}}{2} Y^2 + \frac{ \tilde{m}_{HX}}{\sqrt{2}} Y$.
The full $M^2$ matrix may be written as 
\begin{align}
M^2 = \begin{bmatrix}
m_{1}^{Yh} &0&0 \\
0&m_{1}^{G}&0 \\
0&0& m_{ff}
\end{bmatrix}.
\end{align}
Due to the assumption that at the inflationary epoch there is no quark condensate, there are no scalar-fermion entries in this matrix. 
Moreover, because of the assumption that there is no Higgs vev during the inflation, the contributions from the fermionic sector described by $m_{ff}$ are proportional to the purely 
gravitational terms. They could be either absorbed into the definition of the renormalized tree-level constants or are subleading in comparison to the scalar contributions.
The last statement stems from the fact that, as will be explained shortly, the terms most important for the inflationary physics will have coefficients proportional to the 
$\xi_{H/X}$ couplings while the coefficients of appropriate terms from the fermionic sector are given by small numbers without any powers of the $\xi_{H/X}$ couplings.
The commutator of the derivatives in the scalar sector is equal to zero. 
Having in mind the expressions written above and using formula (\ref{G1_gen}), we may write relevant terms of the one-loop part of the effective action as 
\begin{align}
\label{one-loop_pot}
\Gamma^{(1)} &=  \int \sqrt{-g}\ d^4 x\; 
\frac{1}{64 \pi^2} \text{tr} \bigg \{
- \left ( m_{1}^{G} \right )^2 \ln \left ( \frac{m_1^{G}}{\mu^2} \right ) + \frac{3}{2} \left ( m_1^{G} \right )^2 +\nonumber \\ 
&- \left ( m_{1}^{Yh} \right )^2 \ln \left ( \frac{m_1^{Yh}}{\mu^2} \right ) 
+ \frac{3}{2} \left ( m_1^{Yh} \right )^2  + \nonumber \\
&- \frac{2}{180} \left( -  R_{\mu \nu} R^{\mu \nu} 
+  R_{\alpha \beta \mu \nu} R^{\alpha \beta \mu \nu}
+ \square R  \right) 
\left [ \ln \left ( \frac{m_1^{Yh}}{\mu^2} \right ) + \ln \left ( \frac{m_1^{G}}{\mu^2} \right ) 
 \right ] + \nonumber \\
& - \frac{1}{3} \square m_{1}^{Yh} \ln \left ( \frac{m_1^{Yh}}{\mu^2} \right )
 - \frac{1}{3} \square m_{1}^{G} \ln \left ( \frac{m_1^{G}}{\mu^2} \right )
\bigg \},
\end{align}
where $\text{tr}$ stands for the ordinary matrix trace.
The same type of an expression could be also derived from the renormalization group arguments, as explained in \cite{Buchbinder_Odintsov_Shapiro_1992}. 
Taking into account the hierarchy of the scales elucidated previously ($-R > m_{X}^2, m_{X}^2 \gg  |m_{H}^2|, Y = 0, h =0 $) 
we may rewrite the mass matrices in the following forms:
\begin{align}
\tilde{M}_{G}^2 \approx M_{hh}^2 \approx \left ( \frac{1}{6} - \xi_H \right )R, \\
M_{YY}^2 \approx m_{X}^2 + \left ( \frac{1}{6} - \xi_X \right )R.
\end{align}
Then the one-loop corrections to the effective action for this system are
\begin{align}
\Gamma^{(1)} = \frac{1}{64 \pi^2} &\int \sqrt{-g}\ d^4 x \left \{
\frac{1}{3} \square M_{YY}^2 \ln \left ( \frac{M_{YY}^2}{\mu^2} \right ) +
\frac{4}{3} \square M_{G}^2 \ln \left ( \frac{M_{G}^2}{\mu^2} \right ) + \right. \nonumber \\
&\left. + \frac{3}{2} M_{YY}^4 + 6 M_{G}^4 - M_{YY}^4 \ln \left ( \frac{M_{YY}^2}{\mu^2} \right ) 
- 4 M_{G}^4 \ln \left ( \frac{M_{G}^2}{\mu^2} \right ) + \right. \nonumber \\
&\left. - \frac{1}{90} \left( -R_{\alpha \beta}R^{\alpha \beta} + R_{\alpha \beta \mu \nu}R^{\alpha \beta \mu \nu}  
+ \square R \right)
\left [ \ln \left ( \frac{M_{YY}^2}{\mu^2} \right ) + 4 \ln \left ( \frac{M_{G}^2}{\mu^2} \right )  \right ]
\right \}.
\end{align}
One important observation is that the formula we presented above is an approximate one. Namely, we discarded all terms that are of the 
orders $\mathcal{O} \left (\frac{\mathcal{R}^3}{M^2} \right ) $, 
$ \mathcal{O} \left ( \frac{ \nabla \nabla}{M^2} \right ) $
and higher. Having this in mind we may observe that after integration by parts the first two terms 
give us $\square M^2 \ln M^2 \sim \frac{ \nabla M^2 \nabla M^2}{M^2}$ and therefore they may be discarded.
The same is true for the $\square R \ln M^2$ terms.
For the inflationary era our large mass parameter is actually the curvature itself, i.e., 
$M_{YY}^2 \sim  \left ( \frac{1}{6} - \xi_X   \right ) R$ and 
$M_{G}^2 \sim   \left ( \frac{1}{6} - \xi_H  \right ) R$. From these formulas we may see
that as long as $| \left (\frac{1}{6} - \xi \right )|$ is of the order $\mathcal{O} (10^2)$
or bigger during the inflation period our approximation is a good one. 
This leads to the following formula: 
\begin{align}
\Gamma^{(1)} &= \frac{1}{64 \pi^2} \int \sqrt{-g}\ d^4 x \left \{
\frac{3}{2} \left [ m_{X}^2 + \left ( \frac{1}{6} - \xi_X \right )R \right ]^2 + 
6 \left [\left ( \frac{1}{6} - \xi_H \right )R \right ]^2 + \right. \nonumber \\
&\left. - \left [ m_{X}^2 + \left ( \frac{1}{6} - \xi_X \right )R \right ]^2 \ln \left ( \frac{ m_{X}^2 + 
\left ( \frac{1}{6} - \xi_X \right )R  }{\mu^2} \right ) + \right. \nonumber \\
&\left.- 4 \left [ \left ( \frac{1}{6} - \xi_H \right )R \right ]^2 
\ln \left ( \frac{ \left ( \frac{1}{6} - \xi_H \right )R  }{\mu^2} \right ) + \right. \nonumber \\
&\left. - \frac{1}{90} \left( -R_{\alpha \beta}R^{\alpha \beta} + R_{\alpha \beta \mu \nu}R^{\alpha \beta \mu \nu} \right)
\ln \left ( \frac{ m_{X}^2 + \left ( \frac{1}{6} - \xi_X \right )R  }{\mu^2} \right ) + \right. \nonumber \\
&\left.  - \frac{4}{90} \left( -R_{\alpha \beta}R^{\alpha \beta} + R_{\alpha \beta \mu \nu}R^{\alpha \beta \mu \nu}  \right) \ln \left ( \frac{ \left ( \frac{1}{6} - \xi_H \right )R }{\mu^2} \right ) 
\right \}.
\end{align}

\subsection{Inflationary setup}

After integrating out the fluctuation of the matter fields we obtained the following form of the gravitational 
action for the inflationary period:
\begin{align}
\label{inflation_R_gen}
S_{inf} &= \int \sqrt{-g}\ d^4 x\; \bigg \{
- \frac{1}{2 \kappa} \left( R + 2 \Lambda \right) + \alpha_1 R_{\mu \nu \rho \sigma}R^{\mu \nu \rho \sigma}
+ \alpha_2 R_{\mu \nu}R^{\mu \nu} + \alpha_3 R^2  + \nonumber \\
&+ \frac{1}{64 \pi^2} \bigg \{
\frac{3}{2} \left [ m_{X}^2 + \left ( \frac{1}{6} - \xi_X \right )R \right ]^2 + 
6 \left [\left ( \frac{1}{6} - \xi_H \right )R \right ]^2 +  \nonumber \\
&- \left [ m_{X}^2 + \left ( \frac{1}{6} - \xi_X \right )R \right ]^2 
\ln \left ( \frac{ m_{X}^2 + \left ( \frac{1}{6} - \xi_X \right )R  }{\mu^2} \right ) + \nonumber \\
&- 4 \left [ \left ( \frac{1}{6} - \xi_H \right )R \right ]^2 
\ln \left ( \frac{ \left ( \frac{1}{6} - \xi_H \right )R  }{\mu^2} \right ) +  \nonumber \\
& - \frac{1}{90} \left( -R_{\alpha \beta}R^{\alpha \beta} + R_{\alpha \beta \mu \nu}R^{\alpha \beta \mu \nu} \right)
 \ln \left ( \frac{ m_{X}^2 + \left ( \frac{1}{6} - \xi_X \right )R  }{\mu^2} \right ) + \nonumber \\
&- \frac{4}{90} \left( -R_{\alpha \beta}R^{\alpha \beta} + R_{\alpha \beta \mu \nu}R^{\alpha \beta \mu \nu} \right)
 \ln \left ( \frac{ \left ( \frac{1}{6} - \xi_H \right )R }{\mu^2} \right ) 
\bigg \}
\bigg \},
\end{align}
where $\kappa = 8 \pi G \equiv \bar{M}_{P}^{-2}$.
As our setup we make a choice of $\alpha_i = 0$, which implies that at the tree-level the gravitational action is represented only 
by the operators of the mass dimension two (the Einstein-Hilbert term plus, possibly, the cosmological constant). 
From the matter sector we have three new parameters which are the mass parameter of the heavy scalar $m_X$ and the nonminimal 
coupling constants for the Higgs field $\xi_H$ and the second scalar $\xi_X$. Furthermore, we assume that 
$\frac{|R_{ei}|}{m_{X}^2} > 1$, where $R_{ei}$ is the value of the Ricci scalar at the end of inflation. 
Taking this into account and assuming that $|R_{ei}| < |R_{bi}|$, where $bi$ stands for the
beginning of the last $50-60$ e-foldings we may ignore $m_{X}^2$ under the logarithms. As far as the nonminimal coupling of the Higgs 
field to gravity is concerned, we have only some mild constraints on its value. Considering this we may assume $\xi_H = \xi_X$ at the 
end of inflation. Taking this and the freedom in the choice of the energy scale $\mu$ into account, we may 
set $\mu^2 = \left ( \frac{1}{6} - \xi_H \right )R = \left ( \frac{1}{6} - \xi_X \right )R$. Such a choice leads to the resummation of the 
logarithms at the end of inflation and due to the fact of rather mild running of the nonminimal coupling 
to the quite good resumation during the last $50-60$ e-foldings. 
After discussing this, let us write the Starobinsky form of the action, disregarding the logarithms for now
\begin{align}
S_{inf} &= \int \sqrt{-g}\ d^4 x \bigg \{ 
- \frac{1}{2 \kappa} \left( R + 2 \Lambda \right) +\nonumber \\
&+ \frac{1}{64 \pi^2} \bigg \{
\frac{3}{2} \left [ m_{X}^2 + \left ( \frac{1}{6} - \xi_X \right )R \right ]^2 + 6 \left [\left ( \frac{1}{6} - \xi_H \right )R \right ]^2 
\bigg \}
\bigg \},
\end{align}          
which may be rewritten as
\begin{align}
\label{inflation_R2}
S_{inf} &= \kappa^{-1} \int \sqrt{-g}\ d^4 x \bigg \{ 
- \left [ \frac{1}{2} + \frac{3 m_{X}^2 \kappa \left ( \xi_X - \frac{1}{6} \right )}{64 \pi^2} \right ] R + \nonumber \\
&+ \left [ 
\frac{3 \kappa \left ( \xi_X - \frac{1}{6} \right )^2}{128 \pi^2} 
+ \frac{6 \kappa \left ( \xi_H - \frac{1}{6} \right )^2}{64 \pi^2}
\right ] R^2
\bigg \},
\end{align}
where we also omitted the contribution from the cosmological constant sector.
To simplify the notation let us introduce the following symbols:
\begin{align}
\bar{\xi}_H \equiv \xi_H - \frac{1}{6}, \qquad \bar{\xi}_{X} \equiv \xi_H - \frac{1}{6}, 
\qquad \xi \equiv \frac{\bar{\xi}_X}{\bar{\xi}_H}, 
\end{align}
and define $a = 64 \pi^2$. Taking this into account we could rewrite the Starobinsky-like part 
of our effective action as
\begin{align}
S_{inf} &= \kappa^{-1} \int \sqrt{-g}\ d^4 x \bigg[
- \frac{1}{2} \left( 1 + \frac{3 \kappa m_{X}^2}{a} \bar{\xi}_X \right) R + \frac{\kappa}{a} \left( \frac{3}{2} \bar{\xi}_X^2 + 6 \bar{\xi}_H^2 \right) R^2
\bigg].
\end{align}
Following the standard analysis of the Starobinsky-type inflation we may fix the values of $\xi_X$ and $\xi_H$. 
To this end, we focus on the heavy scalar mass case, namely we assume that $m_{X} \sim 10 \textrm{TeV}$.
This means that $\kappa m_{X}^2 \ll 1$ and can be discarded in the first approximation. On the other hand,
comparing the coefficient in the front of the $R^2$ term with its typical form in the Starobinsky action we get
\begin{align}
\alpha = \frac{1}{2 a} \frac{15}{2} \bar{\xi}^2,
\end{align}
where we used the fact that in our setup $\bar{\xi}_X = \bar{\xi}_H \equiv \bar{\xi}$ 
and took into account the $\kappa^{-1}$ factor in the front of the action. From the above we obtain
\begin{align}
\bar{\xi} = \sqrt{\frac{ 4 \cdot 64 \pi^2 \alpha}{15}}. 
\end{align}
For a successful Starobinsky inflation we need $\alpha \approx 0.97 \cdot 10^9$ 
\cite{STAROBINSKY198099,Netto2016,Calmet_Kuntz_2016}.
This leads us to
\begin{align}
\bar{\xi} \approx 2.02 \cdot 10^5.
\end{align}
 
Let us now return to the problem of the presence of the logarithmic corrections in the effective action. 
Taking them into account and bringing them to the common factor by substitution 
\begin{align}
\label{log_iden}
\ln \left ( \frac{- \bar{\xi}_{X} R}{\mu^2} \right ) = 
\ln \left ( \frac{(- \bar{\xi}_{X} R)}{(- \bar{\xi}_{H} R)} \frac{( - \bar{\xi}_{H} R)}{\mu^2} \right ) =
\ln \xi + \ln \left ( \frac{- \bar{\xi}_{H} R}{\mu^2} \right ) 
\end{align}
we may write our general form of the inflationary action (\ref{inflation_R_gen}) as
\begin{align}
\label{S_inf}
S_{inf} &= \kappa^{-1} \int \sqrt{-g}\ d^4 x \bigg \{
-\frac{1}{2} \bigg [ 1 + \frac{\kappa m_{X}^2}{a}  \bar{\xi}_X
\big( 3 - 2 \ln \xi \big)  \bigg ]R + \nonumber \\
&+ \frac{\kappa}{a} \bigg [ 6 \bar{\xi}_{H}^2 +
\frac{1}{2} \bar{\xi}_{X}^2 \big( 3 - 2 \ln \xi \big) \bigg ]R^2
- \frac{\kappa m_{X}^2}{a} m_{X}^2 \ln \left ( \frac{ - \bar{\xi}_{H} R }{\mu^2} \right ) + \nonumber \\
&+ \frac{2 \kappa  m_{X}^2}{a} \bar{\xi}_{X} R \ln \left ( \frac{ - \bar{\xi}_{H} R }{\mu^2} \right )
- \frac{\kappa}{a} \big( \bar{\xi}_{X}^2 + 4 \bar{\xi}_H^2  \big) R^2 
\ln \left ( \frac{ - \bar{\xi}_{H} R }{\mu^2} \right ) + \nonumber \\
& - \frac{\kappa}{a} \frac{1}{18} \big( \mathcal{K} - R_{\mu \nu} R^{\mu \nu} \big)
\ln \left ( \frac{ - \bar{\xi}_{H} R }{\mu^2} \right )
\bigg \},
\end{align}
where $\mathcal{K} \equiv R_{\alpha \beta \mu \nu}R^{\alpha \beta \mu \nu}$ is the Kretschmann scalar. Using the identity (\ref{log_iden})
beside the term $- \frac{\kappa}{a} \frac{1}{18} \big( \mathcal{K} - R_{\mu \nu} R^{\mu \nu} \big)
\ln \left ( \frac{ - \bar{\xi}_{H} R }{\mu^2} \right )$ we also obtained the $- \frac{\kappa}{a } \frac{\ln \xi}{90} \left( \mathcal{K} - R_{\mu \nu}R^{\mu \nu} \right)$ one.
This finite term can be absorbed into the definition of the renormalized $\alpha_1$ and $\alpha_2$ constants and since we choose $\alpha_1 = \alpha_2 = 0$ is irrelevant for further calculations.

From the obtained form of the inflationary action (\ref{S_inf}) we may see that integrating out 
the matter fields leads not only to the appearance of the Starobinsky-like terms $R + \alpha R^2$ and
usually assumed $R^2 \ln R$ corrections but also to an occurence of terms of the form
$\ln R$, $R \ln R$ and $(\mathcal{K} - R_{\mu \nu} R^{\mu \nu}) \ln R$. The presence of the last of these
terms was not previously discussed in the literature. Moreover, we see that the coefficient of the $R^2$ term 
is controlled by the value of the nonminimal coupling between the scalar field and gravity and could be 
in principle freely chosen to get a correct inflationary prediction. On the other hand, the coefficients
of the $\ln R$ and $R \ln R$ terms are controlled by the mass parameter of the scalar field. 
At this point we want to note that formally there should also be terms of the same form with coefficients 
proportional to the Higgs field mass parameter $m_{H}^2$ but we dropped them on the basis on our hierarchy 
assumption $m_{X}^2 \gg |m_{H}^2|$. On a side note, the term proportional to the 
$\left( \mathcal{K} - R_{\mu \nu} R^{\mu \nu} \right) \ln \left ( \frac{ - \bar{\xi}_{H} R }{\mu^2} \right )$
has a coefficient that is not controlled by any of the matter field couplings. This means that it will be
present for any action containing scalar fields. On the other hand, for the case of the nonminimally coupled scalar
we need $\bar{\xi} \sim 10^5$ for the Starobinsky-like inflation. Hence, we conclude that in the case at hand this term will be much smaller than the $R^2 \ln R$ one.

Having reached the above conclusion, we may write our inflationary action as the $F(R)$ gravity action with 
unusual logarithmic terms 
\begin{align}
\label{SinfR}
S_{inf} &= \frac{1}{2 \kappa} \int \sqrt{-g}\ d^4 x \bigg[
-R + c_{R^2} R^2 
-c_{log} \ln \left ( \frac{- \bar{\xi}_{H} R }{\mu^2} \right ) + \nonumber \\
&+ c_{R log} R \ln \left ( \frac{- \bar{\xi}_{H} R }{\mu^2} \right )
- c_{R^2 log} R^2 \ln \left ( \frac{- \bar{\xi}_{H} R }{\mu^2} \right )
\bigg],
\end{align}
where
\begin{align}
c_{R^2} &= \frac{\kappa}{128 \pi^2} \bigg [ 6 \bar{\xi}_{H}^2 +
\frac{1}{2} \bar{\xi}_{X}^2 \bigg ( 3 - 2 \ln \xi \bigg ) \bigg ], \\
c_{log} &= \frac{\kappa m_{X}^2}{128 \pi^2} m_{X}^2, \\
c_{R log} &= \frac{ \kappa  m_{X}^2}{ 64 \pi^2} \bar{\xi}_{X}, \\
c_{R^2 log} &= \frac{\kappa}{128 \pi^2} \left( \bar{\xi}_{X}^2 + 4 \bar{\xi}_H^2  \right).
\end{align} 
In the above coefficients we restored the loop factor $a = 64 \pi^2$.

\subsection{Inflationary potential after the conformal transformation and running of nonminimal couplings}

\begin{figure}[tbp]
\centering
\includegraphics[width=.85\textwidth]{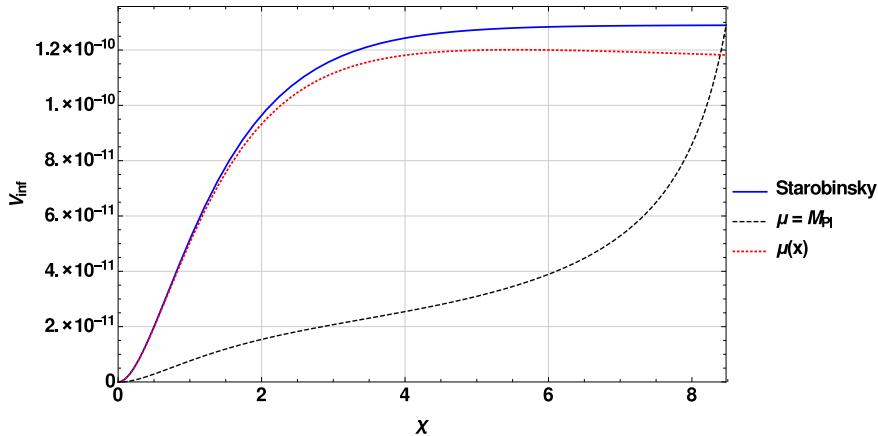}
\caption{The inflationary potential obtained after the conformal transformation for the purely Starobinsky case and for the logarithmically corrected case.
The energy scale was chosen to be either constant or the running one. The nonminimal couplings were constant and set as $\xi_X = \xi_H = 2.02 \cdot 10^5$.}
\label{fig1}
\end{figure}

To discuss the inflationary sector we may introduce an auxiliary scalar field $\chi$. To this end, we represent the inflationary action (\ref{SinfR})
in the form of the $F(R)$ gravity and follow the iterative procedure outlined in 
\cite{BenDayan_Jing_Torabian_Westphal_Zarate_2014}. 
The new scalar field is defined by the relation $e^{\chi} = - \frac{d F(R)}{ d R}$, where the minus sign is
due to the assumed sign convention concerning the Ricci scalar. In this new framework the inflationary potential is given by 
\begin{align}
\label{Vinf}
V_{inf} = \frac{1}{2 \kappa} \frac{R \frac{d F(R)}{ d R} - F(R)}{\left (- \frac{d F(R)}{ d R} \right )^2}.
\end{align} 
From now on, we set $\kappa = 1$. Before we describe the behavior of $V_{inf}$ let us write all the functions relevant for its construction, i.e.,
\begin{align}
\label{F_R}
F(R) &= -R + c_{R^2} R^2 
-c_{log} \ln \left ( \frac{- \bar{\xi}_{H} R }{\mu^2} \right )
+ c_{R log} R \ln \left ( \frac{- \bar{\xi}_{H} R }{\mu^2} \right ) + \nonumber \\
&- c_{R^2 log} R^2 \ln \left ( \frac{- \bar{\xi}_{H} R }{\mu^2} \right ), \\
\frac{d F(R)}{ d R} &= - \left( 1 + \frac{c_{Rlog} \mu^2}{\bar{\xi}_H} \right) + 
\left( 2 c_R + \frac{c_{R^2 log} \mu^2}{\bar{\xi}_H} 
\right)R + \frac{c_{log}}{R}  + \nonumber \\
&+ c_{Rlog} \ln \left ( \frac{- \bar{\xi}_H R}{\mu^2} \right ) 
- 2 c_{R^2 log} R \ln \left ( \frac{- \bar{\xi}_H R}{\mu^2} \right ), \\
e^{\chi} &=  \left( 1 + \frac{c_{Rlog} \mu^2}{\bar{\xi}_H} \right) -
R \bigg[ \left( 2 c_R + \frac{c_{R^2 log} \mu^2}{\bar{\xi}_H} 
\right) + \frac{c_{log}}{R^2} + \nonumber \\
&+ c_{Rlog} \ln \left ( \frac{- \bar{\xi}_H R}{\mu^2} \right ) \frac{1}{R}
- 2 c_{R^2 log} \ln \left ( \frac{- \bar{\xi}_H R}{\mu^2} \right ) \bigg].
\end{align}
In the case of $c_{log} = c_{Rlog} = 0$ the last of the above relations can be solved exactly to obtain $R(\chi)$ and the result is given by
\begin{align}
R = \frac{ e^{\chi} - 1}{ 2 c_{R^2 log} W \left ( - \frac{\bar{\xi}_H \left ( e^{\chi} - 1\right ) }{ 2 c_{R^2 log} \mu^2} e^{- \left ( \frac{c_{R^2}}{ c_{R^2 log}} + \frac{\mu^2}{2 \bar{\xi}_h} \right )} \right )},
\end{align}
where $W(x)$ is the Lambert $W$ function. When $c_{log} = c_{Rlog} \neq 0$ we were unable to find a solution in terms of the known special functions. 
In this case we resorted to the iterative procedure mentioned earlier. For this purpose, we rewrote the equation to be solved as
\begin{align}
e^{\chi} = \left( 1 + \frac{c_{Rlog} \mu^2}{\bar{\xi}_H} \right) -
R \left( 2 c_R + \frac{c_{R^2 log} \mu^2}{\bar{\xi}_H} 
\right) \left [1  + \frac{1}{  2 c_R + \frac{c_{R^2 log} \mu^2}{\bar{\xi}_H} } \frac{f(R)}{R} \right ],
\end{align}
where $f(R) = \frac{c_{log}}{R} +  c_{Rlog} \ln \left ( \frac{- \bar{\xi}_H R}{\mu^2} \right ) 
- 2 c_{R^2 log} R \ln \left ( \frac{- \bar{\xi}_H R}{\mu^2} \right )$.
As the zeroth order approximation to the solution to the above equation we get
\begin{align}
\label{R0x}
R_0 = - \frac{ e^{\chi} - \left( 1 + \frac{c_{Rlog} \mu^2}{\bar{\xi}_H} \right)}{ \left( 2 c_R + \frac{c_{R^2 log} \mu^2}{\bar{\xi}_H} \right)}.
\end{align}
The first order corrected solution is given by
\begin{align}
R_1 = - \frac{ e^{\chi} - \left( 1 + \frac{c_{Rlog} \mu^2}{\bar{\xi}_H} \right) }{ \left( 2 c_R + \frac{c_{R^2 log} \mu^2}{\bar{\xi}_H} \right) 
\Bigg[ 1 + \frac{1}{ \left ( 2 c_R + \frac{c_{R^2 log} \mu^2}{\bar{\xi}_H} \right )} \frac{ f(R_0)}{R_0} \Bigg]},
\end{align}
and the n-th order corrected solution is 
\begin{align}
R_n = - \frac{ e^{\chi} - \left( 1 + \frac{c_{Rlog} \mu^2}{\bar{\xi}_H} \right)}{ \left( 2 c_R + \frac{c_{R^2 log} \mu^2}{\bar{\xi}_H} \right) 
\Bigg[ 1 + \frac{1}{ \left ( 2 c_R + \frac{c_{R^2 log} \mu^2}{\bar{\xi}_H} \right )} \frac{ f(R_{n-1})}{R_{n-1}} \Bigg]}.
\end{align}
For comparison, the purely Starobinsky case is given by putting $c_{R^2 log} =0$ in (\ref{R0x}).
For the purpose of further analysis we used the first order solution $R_1$.
As may be seen from the form of $R_1$ our inflationary potential will depend on the renormalization scale $\mu$.
The comparison of the inflationary potential (given by (\ref{Vinf})) for the Starobinsky case and for the discussed case
for two choices of the running energy scale $\mu$ is presented in figure~\ref{fig1}.
As can be inferred from the plot, choosing $\mu$ as some constant value leads to the 
drastic change in the shape of the potential. Moreover, choosing this scale to lay below the Planck mass leads to the  
discontinuity in the potential which is connected to the presence of the logarithmic terms in $\frac{d F(R)}{ d R}$. This leads us to 
the question of the choice of the running energy scale that allows us to resum the logarithms. In fact, the effective action for the inflationary sector (\ref{SinfR}) was 
already written in the form that made this choice explicit, namely $\mu^2 = - \bar{\xi}_H R(\chi)$, where $R(\chi) \equiv R_0(\chi)$ and is given by
(\ref{R0x}). Due to our sign convention we have $ -R > 0$
and we choose $\bar{\xi}_H$ (and $\bar{\xi}_X$) to be always positive. Figure~\ref{fig1} also shows 
that taking into account the running of the nonminimal couplings leads to an appearance of the runaway direction in the 
inflationary potential.

\begin{figure}[tbp]
\centering
\includegraphics[width=.85\textwidth]{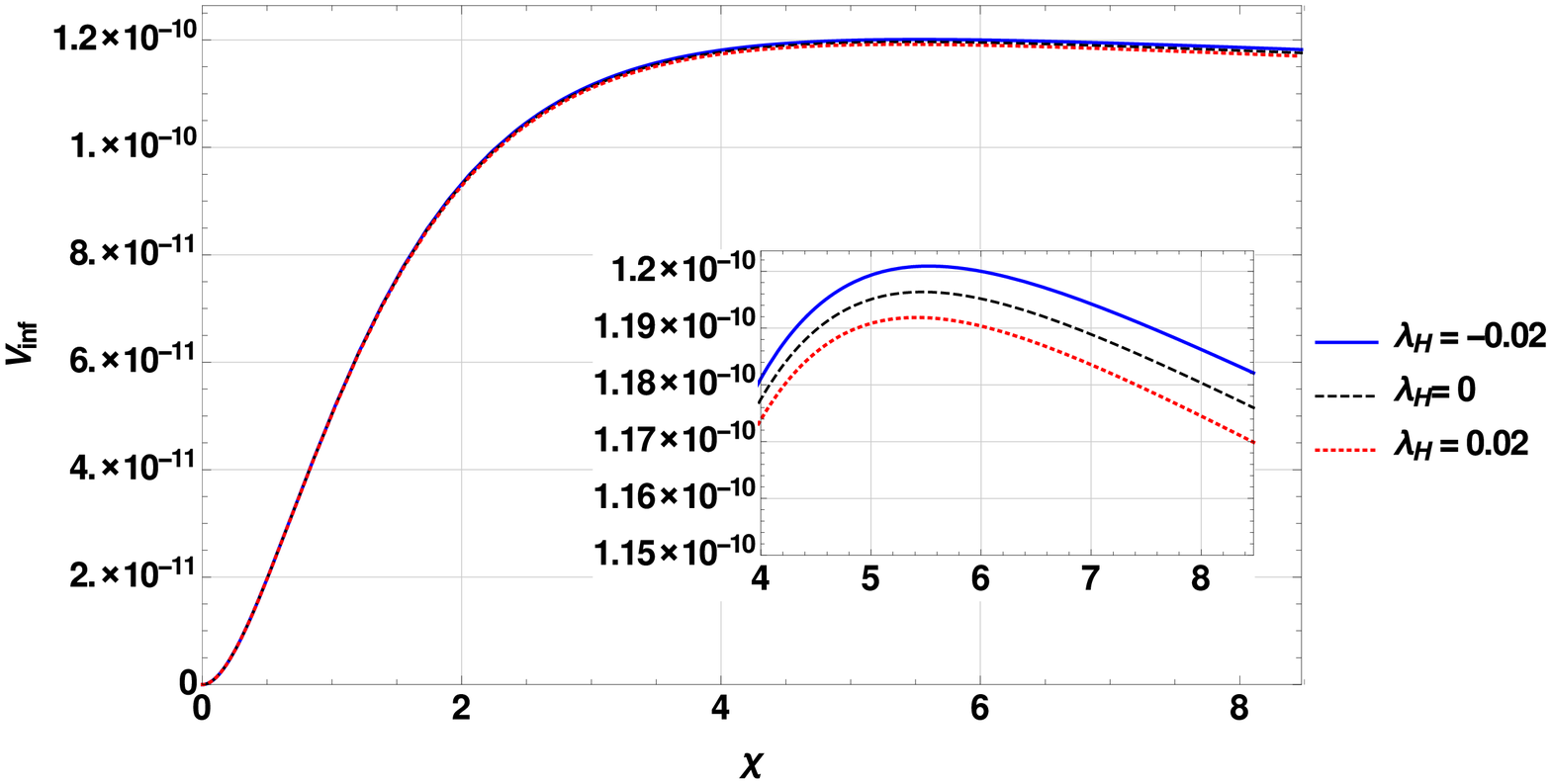}
\caption{The inflationary potential obtained after the conformal transformation. The running energy scale was chosen as $\mu^2 = - \xi_H R(\chi)$.
The relevant matter couplings were chosen as $\lambda_X (\mu_{inf}) = 0$, $\lambda_{HX} (\mu_{inf}) = 0.6$ and $y_{top} (\mu_{inf}) = 0.4$ and were not running.
The initial values of the nonminimal couplings were chosen as $\xi_{H} (\mu_{inf}) = \xi_X (\mu_{inf}) = 2.02 \cdot 10^{5}$, 
where $\mu_{inf} = 10^{16} \textrm{GeV}$.}
\label{fig2}
\end{figure}

\begin{figure}[tbp]
\centering
\includegraphics[width=.85\textwidth]{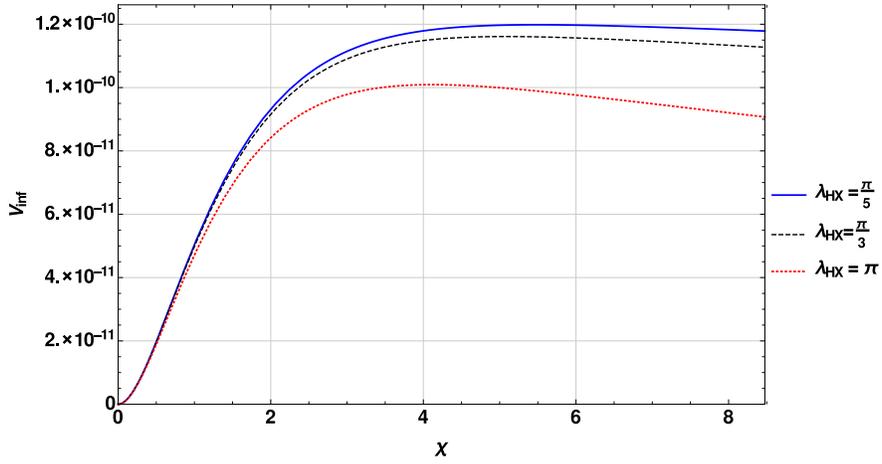}
\caption{The inflationary potential obtained after the conformal transformation. The running energy scale was chosen as $\mu^2 = - \xi_H R(\chi)$.
The relevant matter couplings were chosen as $\lambda_X (\mu_{inf}) = 0$, $\lambda_{H} (\mu_{inf}) = -0.02$ and $y_{top} (\mu_{inf}) = 0.4$ and were not running.
The initial values of the nonminimal couplings were chosen as $\xi_{H} (\mu_{inf}) = \xi_X (\mu_{inf}) = 2.02 \cdot 10^{5}$, 
where $\mu_{inf} = 10^{16} \textrm{GeV}$.}
\label{fig3}
\end{figure}

After resorting to the running energy scale in our analysis of the inflationary potential we should consider the 
running of the nonminimal couplings. For this purpose we need the beta function for these couplings. For the model at hand they were calculated in \cite{Czerwinska_Lalak_Nakonieczny_2015} and are given by 
\begin{align}
\label{beta_xi}
\beta_{\xi_{H}} &= \frac{1}{(4 \pi)^2} \left [ 
3 \lambda_H \left ( \xi_H - \frac{1}{6} \right ) + \lambda_{HX} \left ( \xi_X - \frac{1}{6} \right )
+ 2 y^2 \left ( \xi_H - \frac{1}{6} \right )
\right ], \nonumber \\
\beta_{\xi_{X}} &= \frac{1}{(4 \pi)^2} \left [
6 \lambda_X \left ( \xi_X - \frac{1}{6} \right ) + \lambda_{HX} \left ( \xi_H - \frac{1}{6} \right )
\right ].
\end{align}
As it is clear from the form of $\beta_{\xi_H}$, we disregard the influence of the gauge couplings on the running of $\xi_H$.
Moreover, in our approximation we will treat the scalars quartic coupling and the top Yukawa coupling as constants. This may 
be justified by the fact that in the SM the Higgs quartic coupling and the top Yukawa coupling change very little 
in the energy range $\mu \sim 10^{16} \textrm{GeV} \div \bar{M}_{Pl}$ which is typical for the inflationary period. 
The starting values of the Higgs quartic coupling and the top Yukawa coupling were taken from 
\cite{Degrassi_2012}.
We also assumed that $\lambda_X (\mu_{inf}) \sim 0$ and $\lambda_{HX} (\mu_{inf}) > 0.5$, where $\mu_{inf} \sim 10^{16} \textrm{GeV}$.
The condition for $\lambda_{HX}$ is necessary for our scalar to be a viable dark matter candidate 
\cite{Feng_2015}, 
meanwhile from the perspective of the dark matter phenomenology $\lambda_X$ is unconstrained. As far as the starting values of the 
nonminimal coupling are concerned, we set them as $\bar{\xi}_H (\mu_{inf}) = \bar{\xi}_X (\mu_{inf}) = 2.02 \cdot 10^5$. This was dictated by the 
demand that the purely Starobinsky (only $c_{R^2} \neq 0$) type of the effective action gives us a viable inflationary model. Given the 
values of the SM Higgs quartic coupling and the top Yukawa coupling that can be found for example in 
\cite{Degrassi_2012}, we may infer from (\ref{beta_xi})
that the running of the nonminimal couplings will be controlled by $\lambda_{HX}$.
From this we may infer that the inflationary physics should be most sensitive to a change in this parameter. 
To check this, we plotted the inflationary potential for different values of the Higgs quartic coupling (keeping $\lambda_{HX}$ fixed) in figure \ref{fig2}.
The shape of the potential changes very little even if we change $\lambda_H$ from negative to positive. 
The only visible change is that for the positive Higgs quartic coupling the potential becomes a bit smaller, the difference is below $1\%$. 
In figure \ref{fig3} the inflationary potential for three different choices of $\lambda_{HX}$ is depicted.
Increasing the coupling between the Higgs sector and the dark sector results in a decrease of the overall scale 
of the potential and an increase of the slope in the runaway direction.  

\begin{figure}[tbp]
\centering
\subfloat[The pure Starobinsky case.]{
  \includegraphics[width=.63\textwidth]{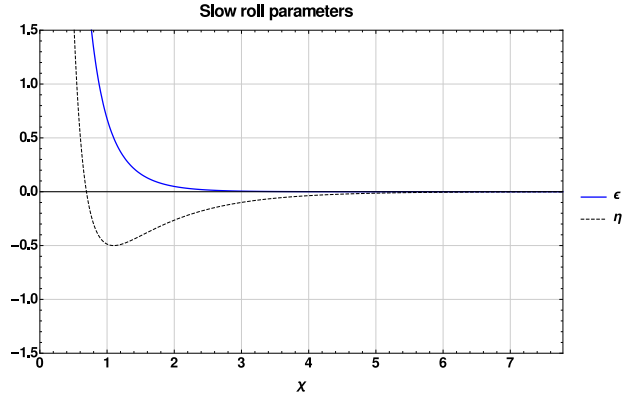}
\label{fig4a}
  }
\hfill
\subfloat[The $\mu = 1$ case and no running of $\xi_{H,X}$.]{
  \includegraphics[width=.63\textwidth]{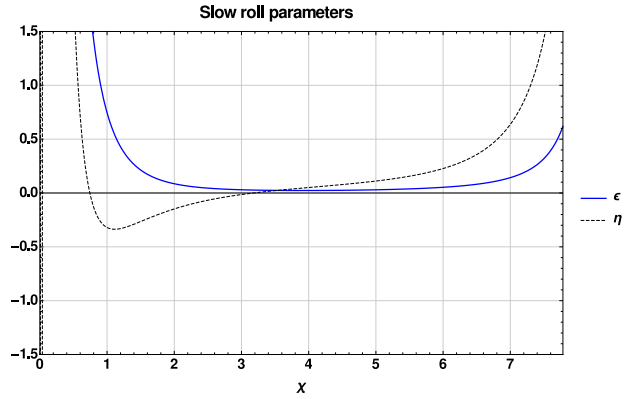}
\label{fig4b}
}
\hfill
\subfloat[The $\mu^2 = - \xi_H R(\chi)$ with the running of $\xi_{H,X}$.]{
  \includegraphics[width=.63\textwidth]{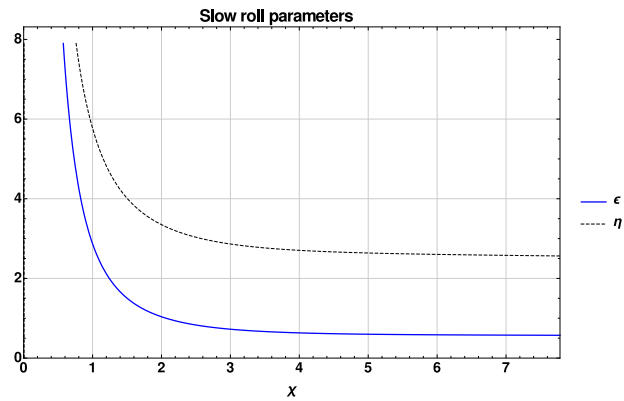}
\label{fig4c}
}
\caption{Slow roll parameters for various potentials presented in figure \ref{fig1}.}
\label{fig4}
\end{figure}

\begin{figure}[tbp]
\centering
\includegraphics[width=.85\textwidth]{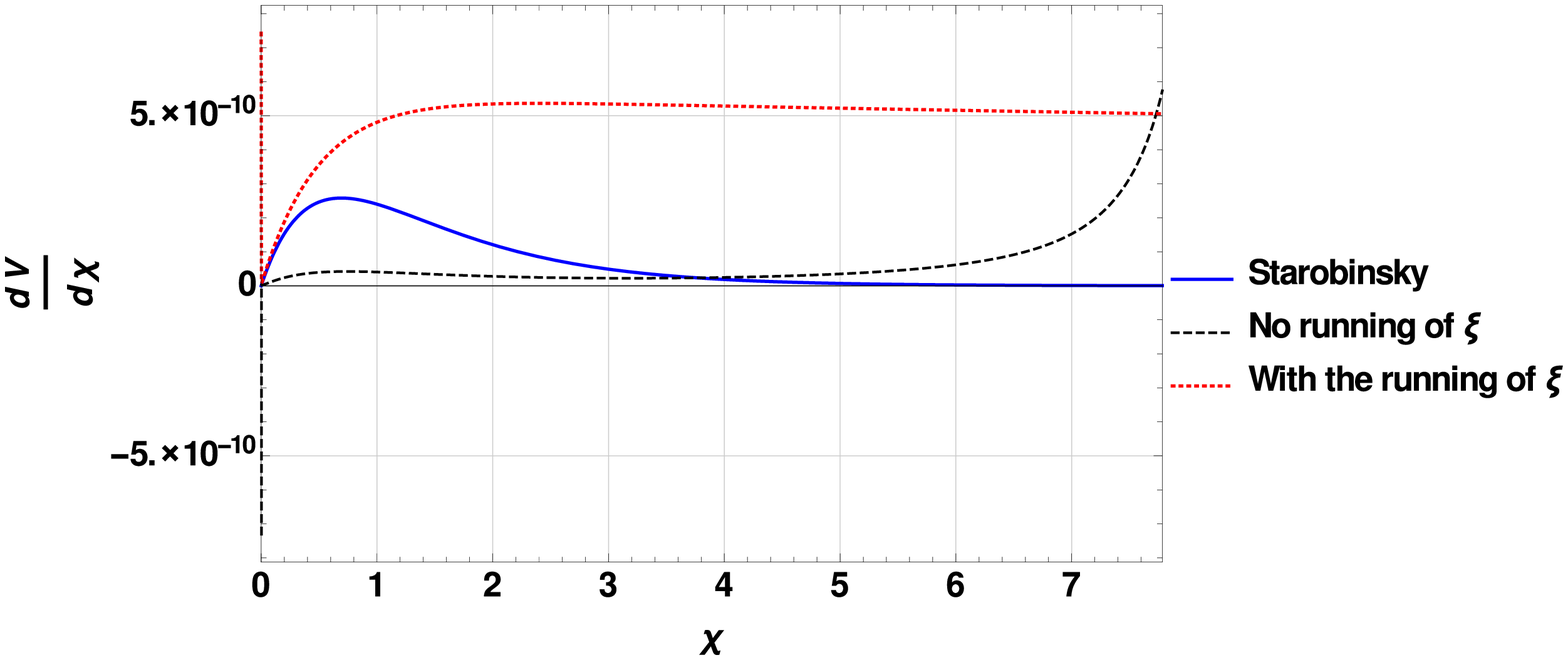}
\caption{Behavior of the first derivative of the potential for the Starobinsky, $\mu =1$ without running and $\mu^2 = -\xi_h R(\chi)$ with the running of $\xi_{H,X}$ cases.}
\label{fig5}
\end{figure}

\begin{figure}[tbp]
\centering
\includegraphics[width=.85\textwidth]{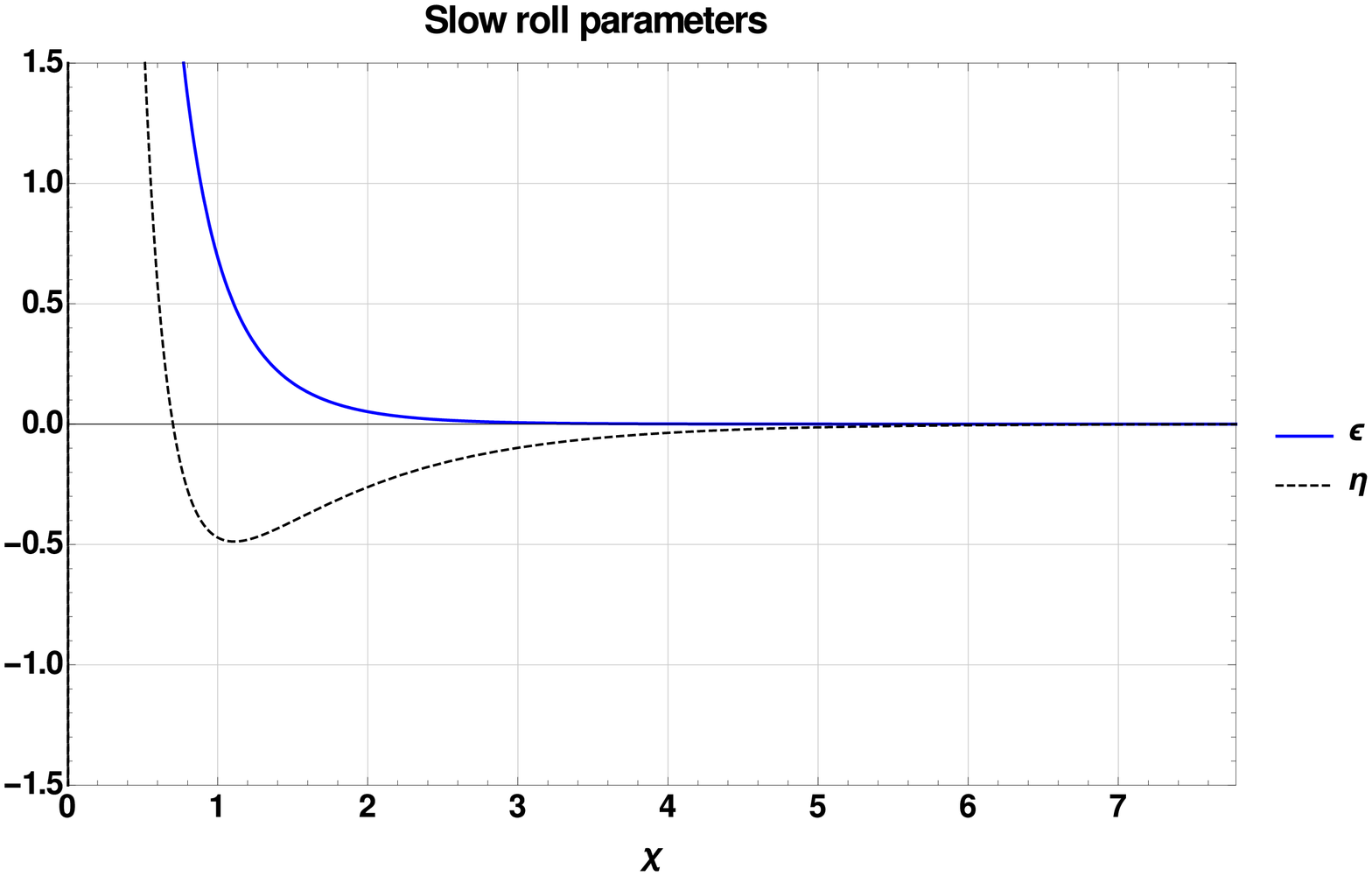}
\caption{Slow roll parameters for $\mu^2 = - \xi_H R(\chi)$ with the running of $\xi_{H,X}$ and the artificially decreased $c_{R^2 \log}$ coefficient.}
\label{fig6}
\end{figure}

The first step in assessing the correctness of the inflationary model is the calculation of the slow roll parameters
from which observables like the spectral tilt and the tensor to scalar ratio can be derived 
\cite{Planck_20_inflation_2015}.
In figure \ref{fig4} we presented the behavior of these parameters for the whole field range.
Remembering that for the slow roll inflationary regime we need $\epsilon < 1$ and $\eta <1$, from figure \ref{fig4b} we may infer that
despite the unusual shape of the potential we still may have good inflationary physics. Moreover, the comparison of figures \ref{fig4a} and \ref{fig4b}
reveals that this model will give us larger tensor to scalar ratio than the purely Starobinsky case (because $r \approx 15 \epsilon$).
On the other hand, from figure \ref{fig4c} we may see that taking into account the running of the energy scale and the running of the 
coupling constants spoils the inflationary model. From the same figure we may infer that in this model there is no slow roll regime. 
This is somewhat unexpected since the potential looks quite similar to the Starobinsky case (the red dotted line in figure \ref{fig1}). 
This difference arises due to the behavior of the derivatives of the potential, which is depicted in figure \ref{fig5}.
The derivative for the running case is slightly bigger than for the Starobinsky case which leads to much
larger $\epsilon$ and this in turn leads to unacceptably large tensor to scalar ratio for this model. 
This is in contrast to the results presented in \cite{Feng_2015}. To comment on that let us point out that there is a large difference 
in the approach to the problem of the quantum corrections to the Starobinsky potential between the one presented in the current paper and in \cite{Feng_2015}.
The authors of the cited paper used phenomenology to fix the coefficients in the front of the $R^2$ and $R^2 \log \left( R \right)$ terms. 
The coefficient of $R^2$ could be fixed from observational data, to be precise it may be fixed by the demand that we get a good inflationary model.
Meanwhile, the coefficient in the front of the second term is usually assumed to be small, in 
the discussed paper the assumption was that the second coefficient is $10^{-2}$ times smaller than 
the first one. 
In our approach both coefficients are generated by the one-loop effects and they are controlled by the tree-level nonminimal couplings between scalars
and gravity. Additionally, there is no evident hierarchy among them (they are of the same order) which we think is ultimately responsible for 
the behavior of the inflationary model obtained by us. As an illustration of this statement we prepared figure \ref{fig6}.
It depicts slow roll parameters calculated for the same case as in figure \ref{fig4c} but with artificially decreased coefficient in the 
front of the $R^2 \log \left( R \right)$ term, namely we took the original coefficient $c_{R^2 \log}$ and divided it by a factor $10^2$.   

To sum up, we may look at the results in a twofold way. Firstly, considering only the Einstein-Hilbert form of the tree-level gravity action 
could the Starobinsky-like inflationary action be generated by the matter loops in curved background? 
Our calculations showed that the answer to this question is negative, since the radiatively generated coefficients of the $R^2$ and $R^2 \log \left( R \right)$ terms 
do not obey necessary hierarchy. 
Secondly, with the inspiration of phenomenology we may introduce the tree-level term proportional to $R^2$, 
with a coefficient fixed by the demand that we obtain a valid inflationary model. For the same reason we may demand that renormalized 
coefficients $\alpha_1 = \alpha_2 = 0$ (there are no $\mathcal{K}$ and $R_{\mu \nu} R^{\mu \nu}$ terms). Now the question is whether the 
one-loop quantum correction from the matter sector in the classical curved spacetime background could destroy this inflationary model. 
In this context we have shown by direct calculations of the one-loop induced coefficients of the terms $R^2$ and $R^2 \log \left( R \right)$
that this may happen if nonminimal couplings of the scalar sector are big enough, i.e., $\bar{\xi} \sim 10^5$ which falls 
into the range usually assumed by the inflationary models based on the scalars nonminimally coupled to gravity.
Resolution to this problem may be as follows. We may introduce $R^2$ at the tree-level with the correct value of the 
coefficient and treat the requirement that the one-loop effect does not spoil the inflationary model as the new constraint on the value of the 
nonminimal couplings between scalars and gravity. Having this in mind, we may interpret our results as indication 
that allowed value of $\bar{\xi}$ should be smaller than $10^5$, which is much more severe restriction than $2.6 \cdot 10^{15}$ obtained in \cite{Atkins_Calmet_2013}.

\section{Summary}
\label{sec:summary}

The issue of a consistent description of both inflationary and dark matter 
sectors and their coupling to the Standard Model fields in the framework of quantum field theory in curved spacetime has been discussed. 
The dark matter sector was based on the heavy real scalar singlet coupled to the SM Higgs field via a
quartic term. In particular, we focused on the TeV range mass for the dark matter candidate which is still allowed. 
As a model for the inflationary sector we used the Starobinsky-like approach. At this point it is worth to 
note that the renormalization of the one-loop effective action in curved spacetime requires a presence of the $R^2$ term and 
also $R_{\mu \nu} R^{\mu \nu}$ and $\mathcal{K} \equiv R_{\mu \nu \rho \sigma}R^{\mu \nu \rho \sigma}$ terms. Although on the phenomenological ground we may
put the renormalized coefficients of the last two terms as equal to zero, these terms reappear on the level of 
loop corrections, see (\ref{inflation_R_gen}). The requirement of the renormalizability of the one-loop effective action 
forces us to consider a nonminimal coupling between scalars (the dark matter candidate $\xi_X$ and the Higgs doublet $\xi_H$)
and gravity which turns out to play a crucial role in the inflationary setup described by us. 

Using the heat kernel approach to the effective action we showed by direct calculations that the high curvature/energy 
regime of our theory is described by the $F(R)$ gravity action given by (\ref{F_R}). Moreover, the 
calculations revealed that for the discussed setup the coefficients of the $R_{\mu \nu} R^{\mu \nu}$ and $\mathcal{K}$ terms 
are much smaller than those standing in front of terms proportional to the Ricci scalar. 
To the best of our knowledge, this is the first direct calculation that demonstrates this effect, which justifies 
an approximation in which we discard terms proportional to $R_{\mu \nu} R^{\mu \nu}$ and $\mathcal{K}$.         
Additionally, the same calculation reveals that beside the usual $R^2$ and $R^2 \ln(R)$ terms 
the one-loop induced effective action contains also terms proportional to $\ln(R)$, $R\ln(R)$,
$R_{\mu \nu} R^{\mu \nu} \ln (R)$ and $\mathcal{K} \ln(R)$. Again, the last two terms may be disregarded 
by the argument of the smallness of their coefficients. 
Amongst the remaining terms the two most important are $R^2$ and $R^2 \ln(R)$. 
We showed by direct calculations that if we consider these terms as induced
at the one-loop level by the SM and dark matter fields their coefficients will be controlled by the
nonminimal coupling of the scalars to gravity (provided that $\xi_{X/H} > 10$, usually for the Higgs inflation we have $\xi_H \sim 10^4$).   
Moreover, it turns out that these coefficients are in the present case of the same order. 

Introducing by means of the conformal transformation an auxiliary scalar field $\chi$ we investigated the details of the 
inflationary sector. Further studies led to the conclusion of importance of the proper choice of the 
energy scale introduced in the process of the renormalization. For the fixed scale $\mu^2 = M_{Pl}^2$ we 
found severe deformations of the inflationary potential as depicted in figure \ref{fig1}. Despite somewhat 
unusual shape of this potential this case may still provide a good inflationary model. This conclusion is 
based on the analysis of figure \ref{fig4c} where we presented the slow roll parameters $\epsilon$ and $\eta$.
On the other hand, for the choice of the running energy scale $\mu^2 = - \xi_H R$ we found out that the inflationary 
potential looks quite similar to the Starobinsky case. The new observation is an appearance of the runaway direction 
towards the large field values, which was also noted in 
\cite{BenDayan_Jing_Torabian_Westphal_Zarate_2014}. Before we discuss a consequence of this let us 
turn to the influence of the scalar quartic couplings on the potential. 
Their values in the inflationary regime have an impact on the potential through the running of 
the nonminimal couplings. We observed that the potential possesses only little sensitivity to the exact value of the Higgs quartic coupling 
which is depicted in figure \ref{fig2}. On the other hand, an increase of the value of the coupling between Higgs and the 
dark matter sector ($\lambda_{HX}$) leads to an increase of the slope of the runaway direction (see figure \ref{fig3}).
Let us now return to the inflationary model with the running energy scale. After an analysis of results we found that 
despite the similarity in the shape of the inflationary potential to the Starobinsky case, in our case there is no slow roll regime.
This may be illustrated by figure \ref{fig4c}, which represents slow roll parameters for the studied case. The $\eta$ is always bigger than $1$
and although $\epsilon$ could be smaller than $1$ it will cause too large value of the tensor to scalar ratio. 
In conclusion, the scenario in which $R^2$ and $R^2 \ln(R)$ terms are generated by the matter loops may lead 
to an inconsistent inflation.

Reaching this conclusion we propose to look at the considered problem from another perspective. 
If we introduce in the classical gravity action the $R^2$ term with a value of the coefficient 
appropriate for the Starobinsky-like inflation then by our calculation of the one-loop corrections we can determine the maximal allowed value of the 
nonminimal coupling $\bar{\xi}$. 
The result is that we need $\bar{\xi} \sim 10^4$ or smaller. This could be treated as a new constraint on the possible value of the $\bar{\xi}$.


\section*{Acknowledgements}
{\L}N is grateful to Geraldine Servant for helpful discussions and hospitality during his stay 
at DESY where a part of the research was conducted. {\L}N is also grateful to Micha{\l} Artymowski for stimulating discussions.
{\L}N was supported by the Polish National Science Centre under postdoctoral scholarship \mbox{DEC-2014/12/S/ST2/00332}.
ZL was supported by Polish National Science Centre under research grant DEC-2012/04/A/ST2/00099.



\bibliographystyle{elsarticle-num} 
\bibliography{higgs_infl_v2.bib}

\end{document}